# A Tutorial in Connectome Analysis:
# Topological and Spatial Features of Brain Networks

Running title: Principles of Connectome Analysis

Marcus Kaiser[1, 2, 3]

[1] School of Computing Science, Newcastle University, UK
[2] Institute of Neuroscience, Newcastle University, UK
[3] Department of Brain and Cognitive Sciences, Seoul National University, South Korea

Corresponding author:

Dr Marcus Kaiser
School of Computing Science
Newcastle University
Claremont Tower
Newcastle upon Tyne, NE1 7RU
United Kingdom,

E-Mail: m.kaiser@ncl.ac.uk   Phone: +44 191 222 8161   Fax: +44 191 222 8232

**Abstract**
High-throughput methods for yielding the set of connections in a neural system, the connectome, are now being developed. This tutorial describes ways to analyze the topological and spatial organization of the connectome at the macroscopic level of connectivity between brain regions as well as the microscopic level of connectivity between neurons. We will describe topological features at three different levels: the local scale of individual nodes, the regional scale of sets of nodes, and the global scale of the complete set of nodes in a network. Such features can be used to characterize components of a network and to compare different networks, e.g. the connectome of patients and control subjects for clinical studies. At the global scale, different types of networks can be distinguished and we will describe Erdös-Rényi random, scale-free, small-world, modular, and hierarchical archetypes of networks. Finally, the connectome also has a spatial organization and we describe methods for analyzing wiring lengths of neural systems. As an introduction for new researchers in the field of connectome analysis, we discuss the benefits and limitations of each analysis approach.

Keywords: cortical networks; neural networks; neuronal networks; brain connectivity; connectome; network analysis

# 1. Introduction

The set of connections in neural systems, now called the connectome (Sporns et al., 2005), has been the focus of neuroanatomy for more than a hundred years (His, 1888; Ramón y Cajal, 1892). However, it attracted recent interest due to the increasing availability of network information at the global (Burns and Young, 2000; Felleman and van Essen, 1991; Scannell et al., 1995; Tuch et al., 2003) and local level (Denk and Horstmann, 2004; Lichtman et al., 2008; Seung, 2009; White et al., 1986) as well as the availability of network analysis tools that can elucidate the link between structure and function of neural systems. Within the neuroanatomical network (structural connectivity), the nonlinear dynamics of neurons and neuronal populations result in patterns of statistical dependencies (functional connectivity) and causal interactions (effective connectivity), defining three major modalities of complex neural systems (Sporns et al., 2004). How is the network structure related to its function and what effect does changing network components have (Kaiser, 2007)? Since 1992 (Achacoso and Yamamoto, 1992; Young, 1992), tools from network analysis (Costa et al., 2007b) have been applied to study these questions in neural systems.

What are the benefits of using network analysis in neuroimaging research? First, networks provide an abstraction that can reduce the complexity when dealing with neural networks. Human brains show a large variability in size and surface shape (Van Essen and Drury, 1997). Network analysis, by hiding these features, can help to identify similarities and differences in the organization of neural networks. Second, the overall organization of brain networks has been proven reliable in that features such as small-worldness and modularity, present but varying to some degree, could be found in all human brain networks (and other species, too). Third, using the same frame of reference, given by the identity of network nodes as representing brain regions, both comparisons between subjects as well as comparisons of different kinds of networks (e.g. structural versus functional) are feasible (Rubinov and Sporns, 2010).

The analysis of networks originated from the mathematical field of graph theory (Diestel, 1997) later leading to percolation theory (Stauffer and Aharony, 2003) or social network analysis (Wasserman and Faust, 1994). In 1736, Leonhard Euler worked on the problem of crossing all bridges over the river Pregel in Königsberg (now Kaliningrad) exactly once and returning to the origin, a path now called an Euler tour. These and other problems can be studied by using graph representations. Graphs are sets of nodes and edges. Edges can either be undirected going in both directions or directed (arcs or arrows) in that one can go from one node to the other but not in the reverse direction. A *path* is a walk through the graph where each node is only visited once. A *cycle* is a closed walk meaning a path that returns back to the first node. A graph could also contain *loops* that are edges that connect a node to itself; however, for analysis purposes we only observe simple graphs without loops. In engineering, graphs are called networks if there is a source and sink of flow in the system and a capacity for flow through each edge (e.g. flow of water or electricity). However, following conventions in the field of network science, we will denote all brain connectivity graphs as networks.

For brain networks, nodes could be neurons or cortical areas and edges could be axons or fibre tracts. Thus, edges could refer to the *structural connectivity* of a neural network. Alternatively, edges could signify correlations between the activity patterns of nodes forming *functional connectivity*. Finally, a directed edge between two nodes could exist if activity in one node modulates activity in the other node forming *effective connectivity* (Sporns et al., 2004). Network representations are an abstract way to look at neural systems. Among the factors missing from network models of nodes, say brain areas, are the location, the size, and functional properties of the nodes. In contrast, geographical or spatial networks also give information about the spatial location of a node. Two- or three-dimensional Euclidean coordinates in a metric space indicate the location of neurons or areas. However, location can also be non-metric where the distance between two nodes has ordinal values (e.g. location of proteins given by a reaction compartment within a cell).

The application of network analysis identified several changes during aging and disease that can form biomarkers for clinical applications. During aging, for example, functional connectivity to neighbors at the local level and other brain areas at the global level is reduced particularly affecting frontal and temporal cortical and subcortical regions (Achard and Bullmore, 2007). In schizophrenia, such small-world features were also altered regarding functional connectivity in EEG and fMRI (Micheloyannis et al., 2006; Skudlarski et al., 2010) and structural connectivity using diffusion tensor imaging (van den Heuvel et al., 2010). Alzheimer's disease patients show a link between highly-connected nodes in functional networks and high amyloid-β deposition (Buckner et al., 2009) and abnormal small-world functional connectivity both in EEG (Stam et al., 2007) and in fMRI (He et al., 2008). Functional connectivity for epilepsy patients is enhanced in EEG (Bettus et al., 2008) and shows altered modular organization in MEG (Chavez et al., 2010) whereas DTI structural connectivity showed reduced fractional anisotropy both adjacent to and further apart from cortical lesions in patients with partial intractable epilepsy (Dumas de la Roque et al., 2004). For healthy subjects, the number of steps to go from one node in the fMRI functional network to another was linked to the IQ of that subject (van den Heuvel et al., 2009).

Network analysis techniques can be applied to the analysis of brain connectivity and we will discuss structural connectivity as an example. Using neuroanatomical or neuroimaging techniques, it can be tested which nodes of a network are connected, i.e. whether projections in one or both directions exist between a pair of nodes (Figure 1A). How can information about brain connectivity be represented? If a projection between two nodes is found, the value '1' is entered in the adjacency matrix; the value '0' defines absent connections or cases where the existence of connections was not tested (Figure 1B). The memory demands for storing the matrix can be prohibitive for large networks as $N^2$ elements are stored for a network of $N$ nodes. As most neuronal networks are sparse, storing only information about existing edges can save storage space. Using a list of edges, the adjacency list (Figure 1C), stores each edge in one row listing the source node, the target node, and—for networks with variable connection weight—the strength of a connection.

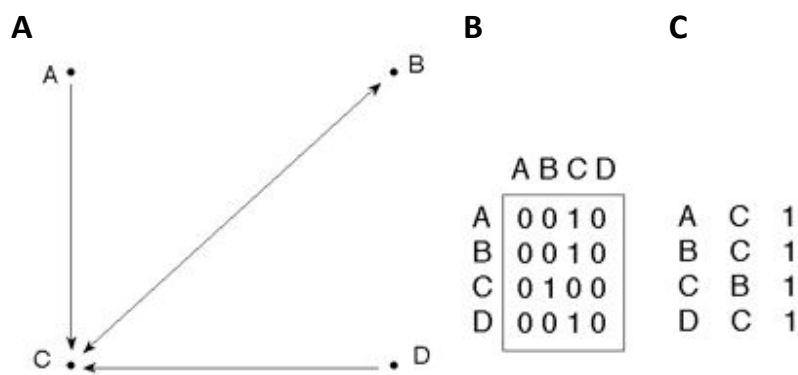

**Fig. 1.** Representations of networks. (A) Directed graph with two directed edges or arcs (A→C and D→C), and one undirected edge being equivalent to a pair of directed edges in both directions (B↔C). (B) The same graph can be represented in a computer using an adjacency matrix where a value of 1 denotes the existence of an edge and 0 the absence of an edge. In this example, rows show outgoing connections of a node and columns show incoming connections. (C) Sparse matrices (few edges) can also be represented as adjacency lists to save memory. Each edge is represented by the source node, the target node, and the weight of the edge (here: uniform value of 1).

In this tutorial, we will neither focus on listing tools for network analysis (Costa et al., 2007b) such as the Brain Connectivity Toolbox nor review connectome analysis results (Bullmore and Sporns, 2009; Sporns et al., 2004). Instead, we will introduce concepts of network analysis to new researchers in the field of brain connectivity. A previous article (Rubinov and Sporns, 2010) provided a classification and discussion of topological network measures relevant to neuroscience. The tutorial here will cover features of the topological but also the spatial organization of neural systems. The spatial location of neurons or regions in three dimensions and the delays for transmitting information over a distance are properties unlike those found in many network models. However, spatial networks with delays for propagation are frequent in real-world artificial and social networks, for example, for epidemic spreading (Hufnagel et al., 2004; Marcelino and Kaiser, 2009; May and Lloyd, 2001), transportation networks (Guimerà and Amaral, 2004; Kaiser and Hilgetag, 2004b), or the Internet (Vázquez et al., 2002; Waxman, 1988; Yook et al., 2002). We discuss the benefits and limits of individual analysis methods. In addition, we will show how to interpret measurements and outline the common pitfalls and misconceptions in the field. This review, for the first time, also introduces the concepts of single node motifs extending notions of hub nodes and provides a classification of cluster detection algorithms. We divide this review into six parts: (remainder of this section) yielding brain networks and the role of node and edge definitions, (section 2) features of individual nodes, (section 3) features of whole networks, (section 4) groups within networks, (section 5) types of networks, and (section 6) spatial network features. Terms, which are also defined in the glossary at the end, are set in *italics*.

## 1.1 Workflow for brain connectivity analysis

How can one get information about brain connectivity between regions, the macroscopic connectome (Akil et al., 2011)? The classical way to find out about structural connectivity is to inject dyes into a brain region. The dye is then taken up by dendrites and cell bodies and travels within a neuron either in an anterograde (from soma to synapse) or a retrograde (from synapse to soma) direction. Typical dyes are Horseradish peroxidase (HRP), fluorescent microspheres, Phaseolus vulgaris-leucoagglutinin (PHA-L) method, Fluoro-Gold, Cholera B-toxin, DiI, and tritiated amino acids. Allowing some time for the tracers to travel, which could be several weeks for the large human brain, the neural tissue can be sliced up and dyes can indicate the origin and target of cortical fiber tracts. Whereas this approach yields high-resolution information about structural connectivity it is an invasive technique usually unsuitable for human subjects (however, there are some post-mortem studies). In the following, we will therefore present non-invasive neuroimaging solutions to yield structural and functional connectivity.

The workflow for yielding human connectivity data starts with anatomical magnetic resonance imaging (MRI) scans with high resolution (Figure 2). These scans are later used to register the location of brain regions. For establishing functional connectivity, a time series of brain activity in different voxels or regions can be derived. The correlation between the time series of different voxels or, using aggregated measures, brain regions can be detected and represented as a correlation matrix (value ranging from -1 to 1). This matrix can either directly be interpreted as a weighted network or it can be transformed into a binary matrix in that only values above a threshold lead to a network connection.

For establishing structural connectivity, diffusion tensor imaging (DTI) or diffusion spectrum imaging (DSI) can be applied. Using deterministic tracking, for example, the number of streamlines between brain regions can be represented in a matrix. For probabilistic tracking, matrix elements would represent the probability to reach a target node starting from a source node. In both cases, the weighted matrix can either be analyzed directly or it can be thresholded so that connections are only formed if a minimum number of streamlines or a minimum probability has been reached.

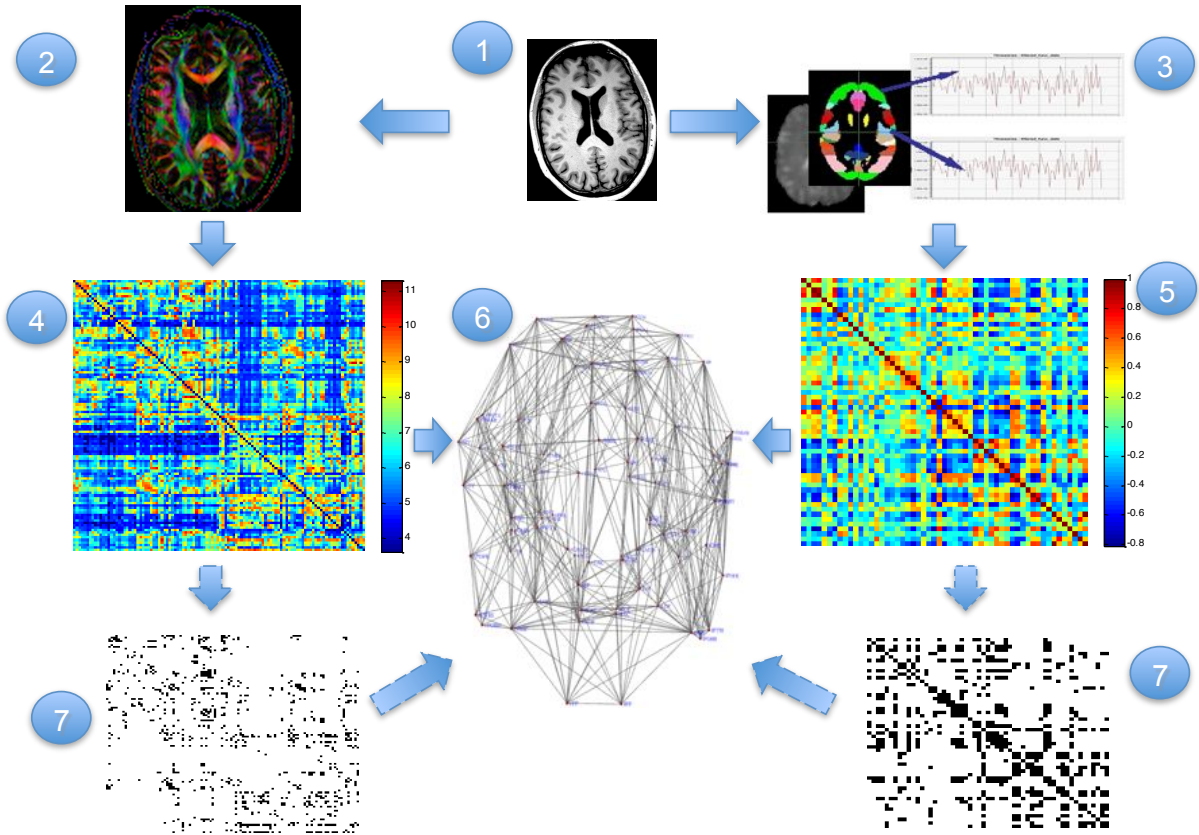

**Fig. 2.** Workflow for structural and functional connectivity analysis. High-resolution anatomical MRI scans of each subject are used as references for further measurements (1). For establishing functional connectivity, a time series of brain activity in different voxels or regions can be derived (3). The correlation between the time series of different voxels or, using aggregated measures, brain regions can be detected and represented as a correlation matrix (5). This matrix can either directly be interpreted as a weighted network (6) or it can be binarized in that only values above a threshold lead to a network connection (7). For establishing structural connectivity, diffusion tensor imaging or diffusion spectrum imaging can be applied (2). Using deterministic tracking, for example, the number of streamlines between brain regions can be represented in a matrix (4). This weighted matrix can either be analyzed directly (6) or be tresholded so that connections are only formed if a minimum number of streamlines has been reached (7).

## 1.2 Role of node and edge definitions

The choice of nodes and edges can be influenced by the anatomical parcellation schemes and measures for determining connectivity (Rubinov and Sporns, 2010). This choice must be carefully considered as different choices might not only change the topology by removing or adding a few nodes or connections but might alter the local and global network features that will be discussed in the following sections.

Parcellations of the brain into different nodes should be (a) non-overlapping in that each brain location only belongs to one region, (b) should assign tissue to one node that has similar connections to other parts of the brain, and (c) should only be compared with other networks that use the same parcellation scheme; this is also crucial for comparing structural and functional connectivity in the same subject (Honey et al., 2009).

Connections of a network can be binarized or weighted. Binary connections only report the absence or presence of a connection. Weighted links can also show the strength of a connection. For structural connectivity, weights can indicate the number of fibers between brain regions (e.g. the streamline count of deterministic tracking), the degree of myelination, the probability that a node can be reached from another node (e.g. probabilistic tracking), or the amount of dye travelling from one node to another (traditional tract-tracing studies). For functional connectivity, weights can indicate the correlation in the time course of signals of different nodes. For effective connectivity, weighted links represent causal relationships between nodes. Weighted networks can be converted to binarized networks using a threshold in that connections are only established if the weight is above a threshold. Using binarized networks simplifies the calculation and interpretation of many network measures (we focus on binarized networks for this tutorial). On the other hand, the choice of a threshold can be problematic: Using the same threshold can lead to a different number of edges in different networks; on the other hand, comparing binarized networks with the same number of edge means that a different threshold may be used for each network (van Wijk et al., 2010). For binarized networks, it therefore has to be noted for which range of thresholds a phenomenon can be observed. In all cases, loops (connections of a node to itself) and negative weights must be removed prior to applying network measures.

## 2. Local scale – single node features

Networks can be characterized at different levels ranging from properties characterizing a whole network at the global scale to properties of network components at the local scale. Starting from the local scale, components of a network are its nodes and edges. Edges can be weighted, taking continuous (metric) or discrete (ordinal) values indicating the strength of a connection. Alternatively, they could just have binary values with zero for absent and one for existing connections.

Thinking about neural systems, there could also be multiple edges between two nodes, e.g. a fiber bundle connecting two brain regions. However, such multi-graph networks are usually simplified in that the number of fibers is either neglected (binary values) or included in the strength of a connection. In addition to fiber count, one might also think of other properties of connections such as delays for signal propagation or degree of myelination. Whereas such properties likely have significant impact on network function, they are currently not part of the analysis of network topology.

The other component at the local scale is a network node. A node could be a single neuron but, as for edges, could also be an aggregate unit of neurons such as a population or a brain area. The *degree* of a node is the sum of its incoming (afferent) and outgoing (efferent) connections. The number of afferent and efferent connections is also called the *in-degree* and *out-degree*, respectively. When $k_i$ denotes the degree of the node *i* of a network with *N* nodes, the series $(k_1,…, k_N)$ with increasing degrees ($k_i \leq k_{i+1}$) is called the *degree sequence* of the network. Nodes with a high number of connections, i.e. a large degree, are called network hubs. For structural and effective connectivity, the ratio between the in- and out-degree of a node can give information about its function: nodes with predominantly incoming connections can be seen as integrators (convergence) whereas nodes with mainly outgoing connections can be seen as distributors (divergence) or broadcasters of information. These distinctions can be useful when nodes are otherwise similar, e.g. distinguishing different types of network hubs (Sporns et al., 2007).

For undirected networks, every connection between nodes is bi-directional (e.g. for functional networks measuring correlation). For such networks, if a node A is connected to a node B with a bi-directional link, this link is counted as one connection when calculating the degree of the node. Likewise, it does not make sense to distinguish in-degree and out-degree as they will have the same values.

Local measures can also refer to the neighborhood of a node. All nodes that directly project to a node or directly receive projections from that node are called *neighbors* of that node. The connectivity between neighbors is used to assess local clustering. The ratio of the number of existing edges between neighbors and the number of potential connections between neighbors forms the *local* *clustering coefficient* (Figure 3A); a measure of neighborhood connectivity.

The local clustering coefficient for an individual node *i* with $k_i$ neighbors (node degree $k_i$) and $\Gamma_i$ edges between its neighbors is

$$C_i = \frac{\Gamma_i}{k_i(k_i-1)}$$

However, this formula is basically not defined if the number of neighbors $k_i$ becomes zero or one as the denominator becomes zero (Costa et al., 2007b). These cases are usually treated as $C_i = 0$ although some authors also set these values to one (Brandes and Erlebach, 2005). Whereas having nodes with no connections (isolated nodes) or only one connection (leaf nodes) is unlikely for structural connectivity, such nodes might occur for functional and effective connectivity. We will discuss some ways to deal with such nodes when describing aggregate measures.

Another local measure of a single node is how many shortest paths (Figure 3B) are containing that node. The shortest path between two nodes is the length of the path with the lowest-possible number of connections. For counting how many paths pass through a node, the shortest paths between all pairs of nodes are calculated. It is then counted how many shortest paths include a certain node. Note that this measure is listed as a local measure as it is an attribute of a single node even though information of the whole network is used to calculate this measure. For comparing this measure for two nodes of the same network, it does not matter whether the absolute number of shortest paths containing one node (stress centrality) or the relative frequency of how often that node is part of shortest paths (betweenness centrality) is taken. This measure of how frequently a node is part of shortest paths is called *node betweenness*. In a similar way, frequency of shortest paths containing a certain edge is the *edge betweenness* of that edge. There are numerous kinds of centrality (Chapter 3, Brandes and Erlebach, 2005) in addition to measures of how many shortest paths run through network components. Other node centrality measures include, for example, closeness centrality, which is the reciprocal of the total distance of a node to any other node of the network, and degree centrality, which is simply the degree of a node.

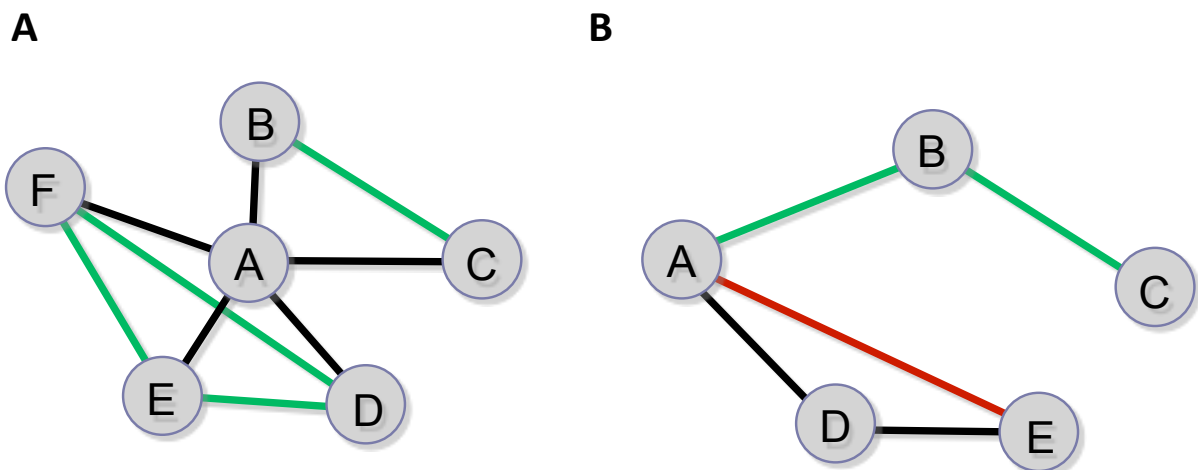

**Fig. 3.** Local and global measures. (A) Local clustering coefficient: All nodes that are connected with node A are neighbors of that node. The local clustering coefficient of node A is the number of connections between neighbors (green edges) divided by the number of all potential connections between its neighbors. In this case, the local clustering coefficient is $C_A=4/10=0.4$ meaning that 40% of connections between neighbors exist. (B) Shortest paths: The shortest path is the path between two nodes with the lowest-possible number of connections in the path. The *path length* of that path is the number of connections that needs to be crossed to go from one node to another. In this example, the length of the shortest path between nodes A and C (A→B→C) is 2 and the length of the shortest path between nodes A and E (A→E) is 1.

## 3. Global scale – aggregate measures

We are now zooming out of a network and observe properties that characterize the network as a whole, leaving out the intermediate regional scale for a moment. Whereas local measures look at properties of individual components, say the primary visual area V1, global measures at the macroscale look at the whole network. This is useful when comparing a given neural network with artificially generated networks called benchmark networks. It is also useful when comparing neural networks from different species or the same species at different levels of organization (area, column, layer). In these cases, the number of nodes and edges as well as their identity (e.g. comparing networks that contain V1 with networks that do not) might differ; however, aggregate measures can still be used to detect changes at the macroscale.

The *edge density*, sometimes called connectivity, of a network is the proportion of connections that exists relative to the number of potential connections of a network. For a directed network with *N* nodes, each node can connect to at most *N*-1 other nodes. Therefore, the edge density of a network with *E* edges and *N* nodes is $d = E / (N (N-1))$. For an undirected network, the edge density becomes $d = E / (2 N (N-1))$; note the factor 2 in the denominator so that any potential undirected edge between two nodes is only counted once and not twice. An edge density of 1, corresponding to a percentage of 100%, would mean that all potential edges exist. In biological networks, however, only a small fraction of potential connections occurs. For the cortico-cortical fiber tract connectivity of the mammalian brain, for example, the edge density ranges between 10 and 30%. For the connectivity between neurons in the nematode *C. elegans*, the edge density is 3.85%.

The edge density gives a first indication how well-connected a network is. However, the time it takes to go from one node to another might still vary considerably depending on the topology of the network. A measure of travelling through a network is the number of connections one has to cross, on average, to go from one node to another. Formally, this *average shortest path* (ASP) of a network with *N* nodes is the average number of edges that has to be crossed on the shortest path from any one node to another:

$$ASP = \frac{1}{N(N-1)} \sum_{i,j} d(i,j) \quad with \quad i \neq j$$

where *d(i, j)* is the length of the shortest path between nodes *i* and *j* having as few connections between nodes *i* and *j* as possible. Note that the definition for the *characteristic path length* L is slightly different (Watts, 1999) in that for each node the average shortest path length to any other node is calculated and the *median*, instead of the mean, value over all nodes is returned as *L*.

In practice, in particular for networks with directed edges, there might be several pairs of nodes for which no path exists. In such cases, graph theory would demand setting the distance *d(i,j)* between the two nodes to infinity. However, having only one such pair in a network would give an *ASP* of infinity as well. In practice, such infinite values are excluded that means the average shortest path only takes the average of existing shortest paths between pairs of nodes. Alternatively, a measure called global efficiency can be used (Achard and Bullmore, 2007; Latora and Marchiori, 2001). Global efficiency uses a sum of the inverse of the distance *L* so that non-existing paths, leading to infinite distance, contribute a zero value to the sum:

$$E_{global} = \frac{1}{N(N-1)} \sum_{i \neq j} \frac{1}{L_{i,j}}$$

where $L_{i,j}$ is the length of the shortest path between nodes *i* and *j* and *N* is the number of nodes.

Another aggregate measure based on the local features of individual nodes is average neighborhood connectivity – the (global) clustering coefficient. The clustering coefficient is just the average of the local clustering coefficient $C_i$ of all nodes:

$$C_1 = \frac{1}{N} \sum C_i.$$

Alternatively, a more widely used definition of the clustering coefficient (Newman et al., 2001) is

$$C_2 = \frac{\sum \Gamma_i}{\sum \deg_i(\deg_i - 1)}.$$

Note that the first definition $C_1$ runs into problems if one node's local clustering coefficient is undefined (see previous section). A third definition is to describe the clustering coefficient using inverse neighborhood clustering analogous to using inverse shortest path lengths for defining efficiency. Such a measure of (neighborhood) disconnectedness $D$ could be defined as (Kaiser, 2008):

$$D = \frac{1}{N} \sum D_i \quad \text{with} \quad D_i = 1/C_i = \frac{\deg_i(\deg_i - 1)}{\Gamma_i}$$
$$\text{and} \quad D_i = 0 \quad \text{for} \quad \Gamma_i = 0.$$

In relation to the characteristic path length as global efficiency (how well are any two nodes of a network connected), the clustering coefficient can also be called local efficiency (how well are neighbours of a node connected) (Achard and Bullmore, 2007; Latora and Marchiori, 2001). The clustering coefficient ($C_1$ or $C_2$) will increase whenever the edge density increases as a higher probability that any two nodes are connected also means that connections between neighbours are more likely. Therefore, comparing clustering coefficents for networks with different edge densities should either be avoided or they should use a normalized coefficient. Such a normalized coefficient would be the clustering coefficient relative to the clustering coefficient of a random network (the clustering coefficient of a random network is the same as the edge density of the original network). However, normalization does not work if one of the two networks has a much higher edge density: whereas the clustering coefficient of a network with low edge density can be 2-3 times as high as the edge density of that network, the clustering coefficient of a network with, say 60% edge density, can never be 2-3 times as high as the edge density.

### 4. Regional scale – Groups of network nodes

Often, we are interested in an intermediate level of organization going beyond single nodes but not including the whole network. At this scale, we can observe measures for subsets of nodes which share similar connections, e.g. dealing with visual input. Such measures look at the sets of nodes where the connectivity within the set is larger than between the set of nodes and the rest of the network. Such sets of nodes are called clusters, modules, or, following social network analysis, communities.

### 4.1 Clusters

Clusters or modules are parts of a network with many connections (high edge density) within such a part and few connections (low edge density) to the remaining nodes of the network. There are many different algorithms to detect clusters of a network (Girvan and Newman, 2002; Hilgetag et al., 2000b; Palla et al., 2005). As a general rule, algorithms can be distinguished along three features (many other classifications exist). First, algorithms could lead to hierarchical or non-hierarchical solutions. Non-hierarchical solutions just identify different modules that can be displayed in a reordered adjacency matrix or in a circular graph (Figure 3A). Hierarchical solutions not only identify modules but also sub-modules within modules, sub-sub-modules within sub-modules, and so on (Clauset et al., 2008). That means that the modular organization at different hierarchical levels can be observed. Instead of distinct levels, the parcellations into sub-modules can also be shown in form of a dendrogram – a tree where nodes in the same cluster are part of the same branch of the tree (Figure 3B). Note that the hierarchical parcellation of the network critically depends on the threshold that one chooses. Whereas an early threshold, near the root of the tree, might only show the main clusters – just one hierarchical level, a late threshold, close to the leafs of the tree, might result in numerous sub-clusters that are so small that the distinction between clusters becomes weak. Second, algorithms can use a predefined number of clusters which need to be detected or can determine the number of clusters themselves. If the number of clusters is known, algorithms similar to *k*-means, where *k* is the number of clusters, can be used to detect the clusters. However, in many biological applications, the number of clusters is not known beforehand, and algorithms that determine the number during the clustering process are needed. Third, algorithms can lead to overlapping or non-overlapping cluster-classifications of nodes. For non-overlapping algorithms, a node will belong to one and only one cluster of the network. However, this assignment to a cluster can often be ambiguous with only a slightly lower preference for assigning the node to another cluster. For overlapping algorithms, a node can belong to several clusters with different likelihoods. Say a node could belong to clusters A, B, and C with likelihoods of 20%, 30%, and 50%, respectively. Such overlapping cluster memberships for cortical nodes can point to nodes that integrate information from several modules, e.g. from the visual and auditory system.

Here, we just show an example of a non-hierarchical, non-overlapping algorithm where the number of clusters is not known beforehand. To identify clusters, an evolutionary optimization algorithm can be used (Hilgetag et al., 2000b). This approach is based on the goal that areas should be more frequently linked to areas in the same cluster than to areas in different clusters. To achieve this goal, Hilgetag et al. defined a two-component cost function *C* whose weighted sum was minimized:

$$C = w_{attr} \times C_{attr} + w_{rep} \times C_{rep}$$

The components were $C_{attr}$ (attraction component), the number of connections existing between clusters, and $C_{rep}$ (repulsion component), the number of absent connections within clusters, with $w_{attr}$ and $w_{rep}$ as weights for adjusting the influence of each component, respectively (Hilgetag et al., 2000a). The first component can be considered as attracting connections to clusters, as it becomes minimal if no connections between clusters exist. Minimizing the second component, on the other hand, tends to break up clusters, as it can be reduced to zero by an arrangement that consists of completely separate areas (Hilgetag et al., 2000a). Only minimizing both components simultaneously will result in dense connectivity within clusters and few connections between clusters.

As an example, this method can be used for the analysis of structural connectivity using fiber tract data in the cat and the macaque as tested for the primate visual, global primate cortical, and global cat cortical network (Hilgetag et al., 2000a). For the cat structural connectivity, based on invasive tract tracing studies, cortical area groupings largely agreed with functional cortical subdivisions (Hilgetag et al., 2000b): the four observed clusters consisted predominantly of visual, auditory, somatosensory-motor, or frontolimbic areas, respectively. In addition, clusters of the primate visual system corresponded closely to the dorsal and ventral visual streams. In agreement with the idea that structural clusters correspond to functional subdivisions, cluster analyses of semi-functional (neuronographic) connections showed functional processing clusters with broadly similar subdivisions (Stephan et al., 2000).

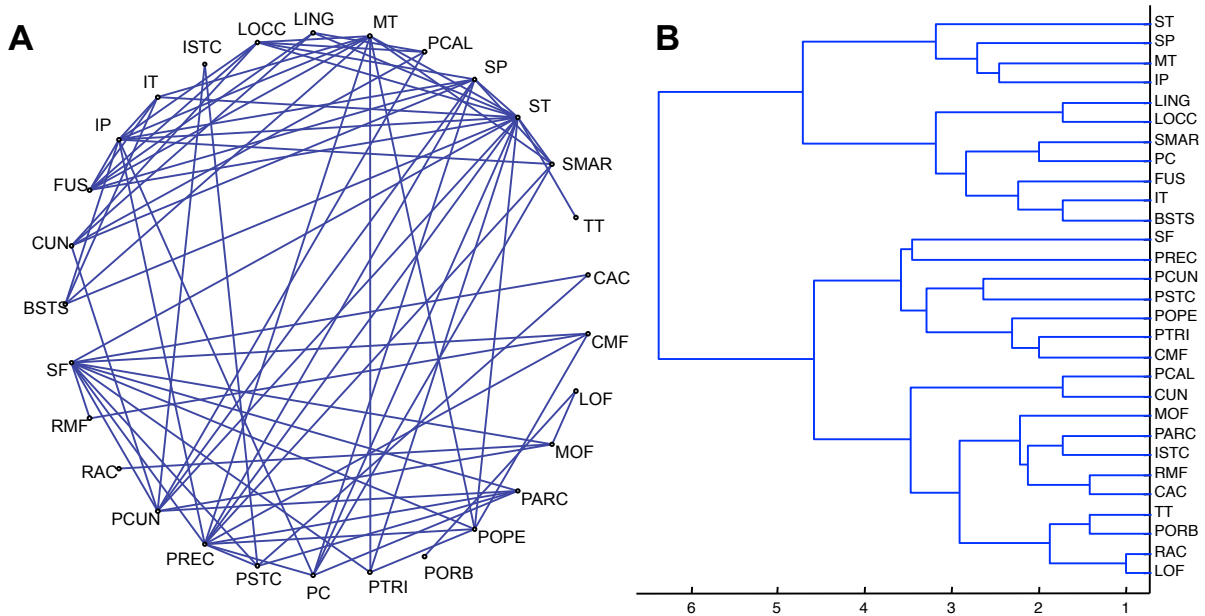

**Fig. 4.** Clusters. (A) Cluster structure of human cortico-cortical connectivity, based on (Hagmann et al., 2008). Cortical areas were arranged around a circle by evolutionary optimization, so that highly inter-linked areas were placed close to each other. Note that nodes in the same cluster, having a high structural similarity, also have a similar function. (B) Dendrogram of the same network using hierarchical clustering. A dendrogram running from the root to the leafs (here: from left to the right) consists of branches connecting objects in the tree. The distance of the branching point on the x-axis is the rescaled distance when clusters are combined.

Figure 4 shows an example of the modular organization of human corticocortical connectivity, based on diffusion spectrum imaging (Hagmann et al., 2008). Cortical areas were arranged around a circle so that highly inter-linked areas were placed close to each other (Figure 4A). Note that nodes in the same cluster, having a high structural similarity, also have a similar function. The cluster architecture of the same network can also be represented by a dendrogram using hierarchical clustering (Figure 4B). A dendrogram running from the left to the right consists of branches connecting objects in the tree. The distance of the branching point on the x-axis is the rescaled distance when clusters are combined.

This is only a brief overview of clustering but there are numerous approaches for detecting network modules. For example, clusters can not only be defined by grouping nodes but also by grouping edges into link communities (Ahn et al., 2010).

### 4.2 Modularity
A measure that has received a lot of attention for topological clusters in recent times is modularity. Modularity ($Q$) is a reflection of the natural segregation within a network (Newman, 2004) and can be a valuable tool in identifying the functional blocks within. Similar to the measure $C$ discussed in the previous sub-section, $Q$ can be used to assess how well a parcellation into non-overlapping modules represents the modular architecture of a network. Given two parcellations into distinct modules for the same network, the parcellation with the higher value of $Q$ would be preferred. So how can the modularity $Q$ be computed? Given a parcellation that assigns to each node $i$ a label $c_i$ identifying to which module the node belongs to, the modularity $Q$ is the difference between the number of edges that lie within a community in the actual network and a random network of the same degree sequence. A high level of topological clustering is reflected in a high value of modularity. The modularity for a directed network is given by (Newman, 2006):

$$Q = \frac{1}{m} \sum_{ij} \left[ a_{ij} - \frac{k_i^{in} k_j^{out}}{m} \right] \delta_{c_i, c_j}$$

where $m$: total number of edges in the network (note that bidirectional links are counted twice); $a_{ij}$ : element of adjacency matrix; $k_i^{in}$ : in-degree of node $i$; $k_j^{out}$ : out-degree of node $j$; $\delta_{c_i c_j}$ :Kronecker delta (only one if nodes $i$ and $j$ are in the same module and zero otherwise); $c_n$ – label of module to which node $n$ belongs to.

This measure can be used as a cost function in cluster algorithms where the aim is to maximize the modularity function $Q$. As for other optimization problems, a range of methods can be used such as genetic algorithms (see previous section), simulated annealing, etc. Note that the modularity measure shows how well a given separation into modules performs. It does not include information about how many modules exist or about their size or overlap (see previous section about clusters on these problems). Using this modularity measure $Q$ also has disadvantages such as limited resolution and a bias towards certain cluster sizes.

Note that this measure differs from the cost function $C$ where the weights $w_{attr}$ and $w_{rep}$ need to be adjusted, in addition to the testing of different cluster membership configurations, to yield a good solution for a particular network. Using the notation of Newman (2006), the cost function $C$ can be re-written as: $C = w_{attr} \sum_{ij} a_{ij}(1 - \delta_{c_i, c_j}) \quad + \quad w_{rep} \sum_{ij} (1 - a_{ij}) \delta_{c_i, c_j}$

### 4.3 Network Motifs

Modules are relatively large structures comprised of tens or hundreds of nodes. Modules are often linked to function in that nodes of the same module tend to have a similar function. However, there could also be smaller subgraphs with only a few nodes that could have a specific function for a network. For a subgraph with only two nodes A and B, there are three ways of how directed edges could exist between them: a connection in one direction, A→B; a connection in the reverse direction, A←B; and a bidirectional connection, A↔B. For sub-graph counting, the case that no connections between the nodes exist is not taken into account. Also, the identity of the nodes is not retained; therefore, A→B and A←B are treated as one pattern. For three nodes, there are already 13 different patterns how directed edges could be distributed (Figure 5A). For a real-world network it is then possible to count how often each potential 2-node or 3-node pattern occurs. If a pattern occurs significantly more often than in a randomly organized network with the same degree distribution, it is called a network *motif*. Dubbed the "building blocks" of complex networks, network motifs mimic the concept of sequence motifs as used in genomics. In a gene sequence, a motif is a recurring subsequence, a pattern that is conjectured to have some functional significance. In a network, a motif is a recurring sub-network conjectured to have some significance.

So when does a pattern occur significantly more often than would be expected for a random organization? To decide this, a set of benchmark networks is generated where the number of nodes and edges is identical but, starting from the original network, edges are rewired while each node maintains its original in-degree and out-degree. Thus, the degree distribution of the network remains unchanged. This means that each node still has the same degree after the rewiring procedure but that additional information, e.g. the cluster architecture, is lost. In the next step, for each benchmark network the number of occurrences of a pattern is determined. Then, the pattern count of the real-world network can be compared with the average pattern count of the benchmark networks; patterns that occur significantly more often in the real-world network than in the benchmark networks are called network motifs.

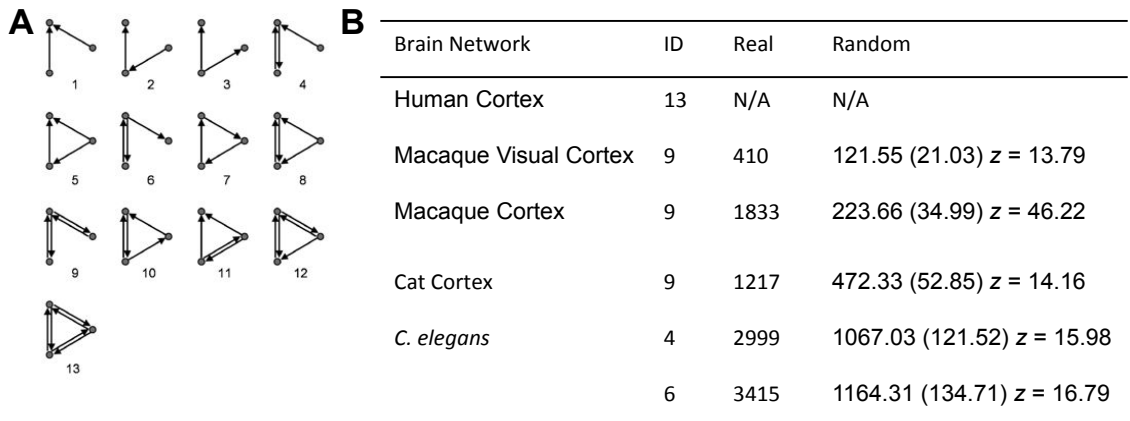

**Fig. 5**. Network motifs. (A) Overview of all 13 possible ways to connect three nodes (three nodes without connections are not considered). (B) Three-node patterns that occur significantly more often in the human (Iturria-Medina et al., 2008), macaque, cat, and *C. elegans* structural connectivity than in rewired networks and are thus network motifs (adapted from (Milo et al., 2002); ID's refer to the numbers in (A)).

The seminal paper by Milo et al. (Milo et al., 2002) gave origin to a multitude of definitions and studies. Network motifs have since been used in the most varied areas. The concept has been applied to networks in domains like protein-protein interactions (PPI) (Alon, 2003; Wuchty et al., 2003), gene transcriptional regulation, food webs, and neural systems (Milo et al., 2002; Sporns and Kötter, 2004). Implementations of motif discovery tools include the original mfinder routine (http://www.weizmann.ac.il/mcb/UriAlon/groupNetworkMotifSW.html ), the faster FANMOD routine (http://theinf1.informatik.uni-jena.de/~wernicke/motifs/index.html (Wernicke and Rasche, 2006)), and a Matlab implementation (http://www.brain-connectivity-toolbox.net/ ). Figure 5A shows all possible patterns for three nodes and Figure 5B shows characteristic network motifs of size 3 for different neural networks (Milo et al., 2002).

Finding these motifs is a computationally hard problem, because fundamentally we match graph patterns with the desired patterns, which leads to the well-known problem of graph isomorphism, with no polynomial time algorithm known (see Box: Runtime complexity). As the size of the motifs gets bigger, the time needed to calculate grows exponentially. Hence, an exhaustive computation of all motifs of a network is typically reduced to very small sizes in order to obtain results in a reasonable amount of time (see (Ribeiro et al., 2009) for a survey of motif detection strategies).

In addition to computational challenges, there are also conceptual problems with motif analysis. The benchmark for counting the number of motifs is a rewired network with the same degree distribution. Even if the degree distribution remains identical, random rewiring removes the topological cluster architecture of the real network that often relates to the underlying spatial clustering of nodes. Taking into account the modular organization and using a rewiring that maintains both the degree distribution and the topological cluster architecture leads to a lower number of network motifs. This occurs as many motifs, such as highly-connected three- or four-node motifs, are frequent when densely connected modules exist. If the rewired network contains such densely connected modules as well, those patterns do not occur significantly more often in the original network. Therefore, only few motifs for the networks considered in (Sporns and Kötter, 2004) remain (Kaiser, unpublished). The same is true for a network with regular connectivity, but without multiple modules: its neighborhood connectivity measured through the clustering coefficient is higher than for the rewired network and thus patterns with dense connectivity arise as network motifs. This is just one example how different features of a network can be strongly correlated. Other examples are the positive correlations between a node degree and its node betweenness, the edge density and the clustering coefficient of a network, and the edge density and the characteristic path length. As mentioned earlier, when comparing networks it is crucial that differences between networks in one network, say the clustering coefficient, are not just caused by differences in another feature, say edge density.

Another problem in applying motif analysis to brain connectivity is that many structural and functional networks yielded by diffusion imaging or time series correlations, respectively, are undirected. For such networks, the number of patterns and therefore potential motifs is significantly reduced: observing 3-node patterns yields 13 potential motifs for directed networks (Figure 5A) but only 3 for undirected networks. This renders motifs less meaningful for undirected networks; however, measurements yielding directed networks might become available in the future.

Network motifs can give information about characteristic patterns of multiple nodes but what about individual nodes that are special for a network? Although a measure for a single node would normally be part of the local scale, we discuss this concept here as it is closely related to multiple-node motifs. Certain singular node-motifs, such as highly connected nodes or hubs, affect spreading phenomena, which makes them important components of the network. More complex compound singular node motifs, which are characterized by multiple features in combination, specify nodes more comprehensively. With this more precise description new kinds of motifs can be formulated (Costa et al., 2009). The algorithm first identifies outlier nodes with features that are significantly different from other nodes in the network. Next, those outlier nodes can be classified into different classes based on their individual features. Finally, the number of nodes that were found for each class gives a fingerprint that is characteristic for each network. As the method does not rely on adjustable parameters, it can be automatically applied to a large number of networks (Echtermeyer et al., 2011). Implementations of this method are available and provide both a graphical user interface and a command line version for batch processing (http://www.biological-networks.org/ ).

## 5. Types of networks

We already mentioned that network measures can be used to compare networks with each other. Often, we are not only interested in how network measures differ but whether the *type* of network differs. Although each neural network has a unique topological and spatial organization, such types or classes of networks can be used for classification and comparison (Figure 6). Such classes are based on global features of the degree distribution and the community organization. The following section shows different types and their characteristic properties. Note that real-world networks, however, might show a combination of different classes, e.g. being modular and small-world.

### 5.1 Random networks

Whereas many networks are generated by a random process, the term random network normally refers to the type of Erdös-Rényi random networks (Erdös and Rényi, 1960). Random networks are generated by establishing each potential connection between nodes with a probability $p$. This probability, for a sufficiently large network, is then equivalent to the edge density of the network; i.e. the connection density. The process of establishing connections resembles flipping a coin where an edge is established with probability $p$ and not established with probability $q = 1 - p$. Therefore, the distribution of node-degrees follows a binomial probability distribution. For large numbers of nodes, the probability $P(k)$ that a node has $k$ connections can be approximated by a Poisson distribution, and hence the term 'exponential degree distribution' is also used (Bollobas, 1985). The distribution can be shown as a histogram where the counts for the different bins are plotted as data points. For the "exponential" degree distribution, $P(k) \sim e^{-k}$, of a random network, points are arranged on a line for a logarithmic plot of $\log(P(k))$.

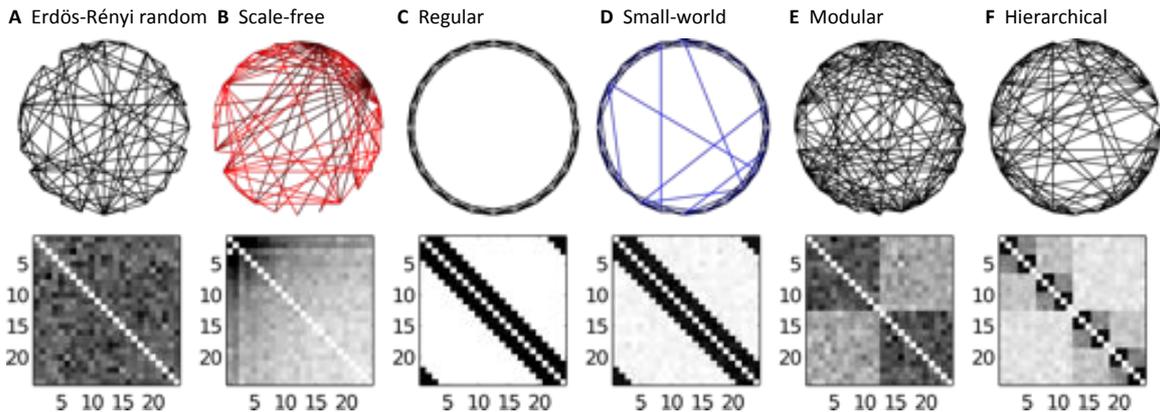

**Fig. 6.** Types of networks. Networks contain 24 nodes and 142 edges. The top panel shows individual networks where nodes are located on a circle. The bottom panel shows the average probability over 100 networks to display a connection in the adjacency matrix: white denotes edges that are always absent whereas black shows edges that are always present. (A) Erdös-Rényi random network. (B) Scale-free network with dark edges (top) indicating highly connected nodes or hubs. (C) Regular or lattice network with high connectivity between neighbors. (D) Small-world network with blue edges (top) representing short-cuts of the network. Note that for the average probability plot (bottom), short-cuts are invisible due the averaging over 100 networks. (E) Modular network with two modules. (F) Hierarchical network with two modules consisting of two sub-modules each. Thus, there are two hierarchical levels of organization.

## 5.2 Scale-free networks

Scale-free networks are characterized by their specific distribution of node degrees. The degree distribution follows a power law where the probability that a node with degree $k$ exists, or the frequency of occurrences for real-world networks, is given by $P(k) \sim k^{-\gamma}$. This is different from the random networks discussed above where the degree distribution follows an exponential distribution, $P(k) \sim e^{-k}$. The exponent $\gamma$ of a power-law degree distribution can vary depending on the network which is studied. For example, for functional connectivity between voxels in human MRI, $\gamma$ was found to be 2.0 (Eguiluz et al., 2005) whereas $\gamma$ was 1.3 for functional connectivity of neurons in the hippocampus determined through calcium imaging (Bonifazi et al., 2009).

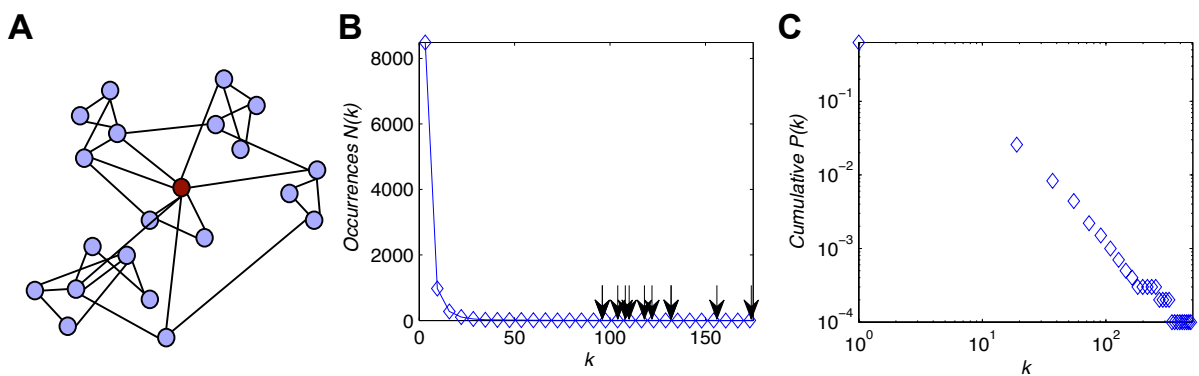

**Fig. 7.** Scale-free networks. (A) Scale-free networks contain highly-connected nodes or hubs (shown in red). (B) Degree distribution of a scale-free network with 10 000 nodes and 20 145 connections. In contrast to random networks with one characteristic scale where all node degrees $k$ are close to the average, scale-free networks can contain nodes with degrees that are several standard deviations away from the average. In this example, there are 13 nodes with degrees that are nine standard deviations away from the average degree of 4 (arrows); the maximum degree is 504 (beyond the figure axes). (C) The cumulative frequency $P(k)$ that a node with degree $k$ occurs in the network follows a power-law leading to a straight line in a bi-logarithmic plot.

Data points for a power-law degree distribution lie on a straight line for a log-log plot of log(P($k$)) against log($k$). To test this power-law relationship, the cumulative distribution $P_c(k) = P(X > k) = 1 - F(k)$ with $F(k) = P(X \leq k)$ where X is the number of connections of a node is plotted (Figure 7C). As the histogram uses the same bin widths, the bins for high-degrees have fewer entries than the bins for nodes with low degree values. Therefore, data points in the histogram will fluctuate more strongly for the tail of the distribution. In addition to the visual inspection of the log-log plot, a statistical analysis is needed to test for a power-law behavior (Clauset et al., 2009). Such an analysis will determine the goodness-of-fit with a power-law distribution and will also compare this result with alternative hypotheses (e.g. exponential or Yule distributions).

There are two potential problems with determining whether a network shows a power-law degree distribution and is thus scale-free: One problem arises when only part of the network is known. For example, one might like to test scale-free properties of neuronal networks but only has connectivity within one column but not with other parts of the brain. In this case, connectivity is only known for a subset or sample of the nodes of the whole network. Using such incomplete sampling, is it still possible to test whether the whole network is scale-free or not? Unfortunately, the amount of unknown or not included connections or nodes might change the shape of the degree distribution (Stumpf et al., 2005); in particular missing nodes could mean that the rarely occurring hubs are not part of the sample resulting in classifying a scale-free network as an Erdös-Rényi random network.

Another problem is networks with a low number of nodes, where the degree distribution only consists of one or two orders of magnitude. Unfortunately, this is the case for regional brain connectivity where networks consist of 100 or less nodes: Structural networks in the macaque, cat, and human within one hemisphere usually consist of around 30-100 nodes. Such a low number of nodes results in power-law fits that are not robust. For such networks, individual outlier nodes with very high connectivity might directly alter the tail of the degree distribution, whereas for larger networks the degrees of several nodes will be binned so that outliers are less influential. However, we can test for scale-free behavior by using indirect measures whose outcome is not altered significantly by one individual node (Kaiser et al., 2007b).

### 5.2.1 Case-Study: Are neural networks scale-free?
Previous studies have shown that functional networks of the human brain, looking at signal correlations between voxels in fMRI, are scale-free (Eguiluz et al., 2005). However, at the gross level of signal correlations between brain regions, it was argued that these functional networks are not scale-free (Achard et al., 2006). We have compared the structural network between cat and macaque brain regions with different types of benchmark networks, including random, scale-free and small-world networks, and found strong indications that the brain connectivity networks share some of their structural properties with scale-free networks. In particular, we compared the effect that the removal of nodes and connections had on the ASP found in the brain connectivity networks and their benchmark counterparts (Kaiser et al., 2007b). So even though the degree distribution cannot be tested, the robustness after simulated lesions is most similar to that of a scale-free network. However, this does not necessarily mean that cortical networks show a power-law degree distribution. Indeed, the structural connectivity is unlikely to be scale-free over more than one order of magnitude. Still, cortical networks contain highly-connected nodes that are unlikely to occur in random networks with the same number of nodes and edges (Kaiser et al., 2007b). These hubs might be the underlying reason for the lesion robustness that was comparable to scale-free networks as described above.

If neural networks show more highly-connected nodes than for random benchmark networks, how could network features such as hubs arise? There are several potential developmental mechanisms that yield brain networks with highly connected nodes. Work in brain evolution suggests that when new functional structures are formed by specialization of phylogenetically older parts, the new structures largely inherit the connectivity pattern of the parent structure (Ebbesson, 1980). This means that the patterns are repeated and small modifications are added during the evolutionary steps that can arise by duplication of existing areas (Krubitzer and Kahn, 2003). Such inheritance of connectivity by copying modules can lead to scale-free metabolic systems (Ravasz et al., 2002). A developmental mechanism for varying the node degree of regions could be the width of the developmental time window for synaptogenesis at different regions (Kaiser and Hilgetag, 2007; Nisbach and Kaiser, 2007). Indeed, *C. elegans* neurons that are generated early during development tend to accumulate more connections and tend to be hubs of the adult network (Varier and Kaiser, 2011).

### 5.3 Small-World networks

Many networks exhibit properties of small-world networks (Watts and Strogatz, 1998). The term small-world refers to experiments in social networks by Stanley Milgram where a person could reach any other person through a relatively short chain of acquaintances, the "six degrees of separation" (Milgram, 1967). However, relatively short does not mean that the average number of connections to cross from one node to another, the characteristic path length, is minimal. Indeed, the path length is usually higher than for Erdös–Rényi random networks with the same number of nodes and edges. However, connectivity between node neighbors, the clustering coefficient, is much higher than for random networks.

So when can a network be considered a small-world network? Unfortunately, there is no clear criterion. In general, to classify a network as small-world, its clustering coefficient should be much higher than the clustering coefficient of Erdös-Rényi random networks. For Erdös-Rényi random networks, the clustering coefficient has the same value as the edge density (connections between neighbors are as likely as any other connections of the network) so edge density might be used for the comparison. In addition, the characteristic path length of the network should be comparable to that of a random network that means slightly but not excessively higher than that value.

A measure to summarize to what extent a network shows features of a small-world network is small-worldness $S = (C / C_{rand}) / (L / L_{rand})$ where $C$ is the clustering coefficient and $L$ is the characteristic path length of an observed network and a random network (Humphries and Gurney, 2008). Note that this measure is useful for comparing small-world networks but not sufficient for determining whether a network is a small-world network or not: a high value of $S$ might also occur for networks with extremely high characteristic path length as long as the clustering coefficient is much higher than for random networks.

#### 5.3.1 Case study: It's a small brain—small-world properties in neural networks

Small-world properties were found on different organizational levels of neural networks: from the tiny nematode *C. elegans* with about 300 neurons (Watts and Strogatz, 1998) over cortical structural connectivity of the cat and the macaque (Hilgetag et al., 2000a; Hilgetag and Kaiser, 2004; Sporns et al., 2000) to human structural (Hagmann et al., 2008) and functional (Achard et al., 2006) connectivity. Whereas the clustering coefficient for the macaque structural connectivity is 49% (16% in random networks), the characteristic path length is comparatively low with 2.2 (2.0 in random networks). Similarly, human structural connectivity between brain regions shows small-world properties with a small-worldness $S$ of 10.6 (Text S2, Hagmann et al., 2008). For human functional connectivity between brain regions, the clustering coefficient is 53% (22% in random networks) and the path length 2.5 (2.3 in random networks) (Achard et al., 2006).

That is, on average only one or two intermediate areas are on the shortest path between two areas. An anatomical basis for small-world features in neural networks is the preference for local short-distance connections with only few long-distance connections (Kaiser et al., 2009). In that way, most neighbors of a node are nearby and therefore have a higher probability to be connected (Kaiser and Hilgetag, 2004a; Kaiser and Hilgetag, 2004b; Nisbach and Kaiser, 2007).

How can small-world networks arise during network development? First, small-world networks, in a method described in the original article (Watts and Strogatz, 1998), could start from a regular (also called lattice) network where neighbors are well connected and rewire connections of nodes by randomly varying the node to which an edge connects to. In this way, edges that do not connect neighbors but distant nodes in a network can be established; these edges are thus called 'short cuts'. If the probability that any edge will be rewired becomes too high, the network turns into a random network. Second, networks with small-world properties can be generated in the reverse way starting with random networks and slowly establishing higher neighborhood clustering (Kaiser, 2008). Such networks can contain isolated nodes and leaf nodes (nodes with only one connection); patterns which are unlikely to arise following the previous approach of starting with regular networks. Third, networks could grow in two- or three-dimensional space with a preference for new nodes to connect to spatially nearby nodes. Such a spatial growth can, in certain parameter regimes, lead to small-world networks (Kaiser and Hilgetag, 2004a; Kaiser and Hilgetag, 2004b; Nisbach and Kaiser, 2007).

Note that a high clustering coefficient does not necessarily mean that a network contains multiple clusters! Indeed, the standard model for generating small-world networks by rewiring regular networks (Watts and Strogatz, 1998) does not lead to multiple clusters. In addition, small-world and scale-free properties are compatible, but not equivalent; a network might be small-world but not scale-free and vice versa.

### 5.4 Modular and hierarchical networks

Two central topological features of brain networks, in particular of the cerebral cortex, are their modular and hierarchical organization. Modular networks consist of multiple clusters (cf. section 4.1 on clusters). If these clusters occur at different levels, a cluster consisting of multiple sub-clusters, sub-clusters consisting of several sub-sub-clusters, and so on, the network can be called a hierarchical modular network. Note that for only one level, the network would be modular but not hierarchical. On the other hand, networks that are hierarchical but not modular seem impossible.

A modular hierarchical organization of cortical architecture and connections is apparent across many scales, from cellular microcircuits in cortical columns (Binzegger et al., 2004; Mountcastle, 1997) at the lowest level, via cortical areas at the intermediate scale, to clusters of highly connected brain regions at the global systems level (Breakspear and Stam, 2005; Hilgetag et al., 2000a; Kaiser et al., 2007a). The precise organization of these features at each level is still unknown, and there exists controversy about the exact organization or existence of modules even at the level of cortical columns (Horton and Adams, 2005; Rakic, 2008). Nonetheless, current data and concepts suggest that at each level of neural organization clusters arise, with denser connectivity within than between the modules. This means that neurons within a column, area or cluster of areas are more frequently linked with each other than with neurons in the rest of the network.

Several potential biological mechanisms for generating hierarchical modular networks have been described. One way is to start with an existing network and generate copies of duplicates of such a network where the copies retain the same internal connectivity as the original network but also establish connections directly to the original networks. Variations of this method can be used to generate hierarchical scale-free networks (Ravasz and Barabási, 2003; Ravasz et al., 2002) and were also thought to lead to cortical connectivity-like networks (Ebbesson, 1980; Krubitzer and Kahn, 2003). For modular networks, time windows during development can lead to multiple clusters where the cluster number, cluster size, and inter-cluster-connectivity is determined by the number, width and overlap of developmental time windows for synaptogenesis (Kaiser and Hilgetag, 2007; Nisbach and Kaiser, 2007).

### 6. Space – the final frontier

The previous sections have looked at topological properties of neural networks but brain networks also have spatial properties in that each node and edge has a three-dimensional location and extension may it be volume for nodes or diameter and trajectory for edges. Given the spatial extent of network components, space is often a limiting factor for the structural organization of neural systems. For example, all-to-all connectivity between all neurons of the brain is impossible given the limited volume available for white matter fiber tracts within the skull. In addition to the feasibility of network topologies, the actual wiring between nodes can also inform us about functional constraints. For example, long-distance connections are costly in that their establishment (material) and maintenance (action potential propagation) uses energy. On the other hand, long-distance connections can form short-cuts that lead to faster information integration and, consequently, accelerated reaction time.

**6.1 Connection lengths**

Each node and edge in neural networks, at least after potential migration during development, has a constant spatial position. Such a spatial layout is far from random but to what extent self-organization or genetic predisposition determines location is still an open question. One first step in observing the spatial organization of a neural network is to look at the lengths of connections. If two nodes are connected, the Euclidean distance between the positions of both nodes can be a lower bound of the length of the connection. Note, that even for cortical fiber tracts this gives a reasonable approximation: for the prefrontal cortex in the macaque only 15% of the connections are strongly curved and dense fibers, in particular, tend to be completely straight (Hilgetag and Barbas, 2006). It is often interesting to observe how many connections go to nearby targets and how many extend over a long distance, potentially linking different components of the neural network. This can be readily observed using a histogram of the connection lengths of a network. These histograms for anatomical connection lengths, ranging from *C. elegans* and rat neuronal to macaque and human cortical connectivity, all show a decay of the frequency over distance: short-distance connections are more frequent than long-distance connections (Figure 8). For these systems, the distribution can best be approximated through a Gamma distribution (see (Kaiser et al., 2009) for details).

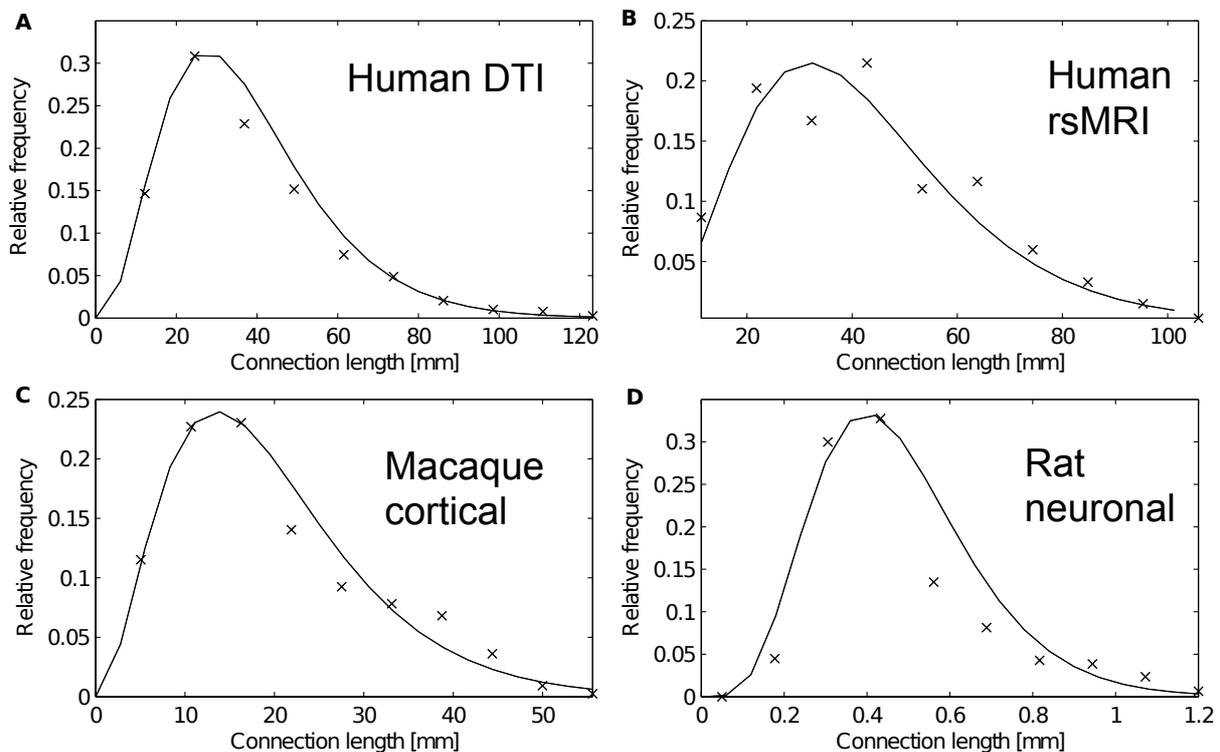

**Fig. 8.** Connection lengths in cortical and neuronal networks. Connection length distributions where the relative counts of a histogram are plotted as data points (x) fitted with a Gamma function (solid line, see Appendix for fits and coefficients). For cortical networks, only connectivity within a hemisphere was considered. Despite different species, levels of organization, and types of connectivity, all distributions show an early peak and a later distance-dependent decay in the frequency of connection. (A) Human diffusion tensor imaging network between 55 brain regions where each existing fiber tract is represented by one (unweighted) network connection. (B) Human resting-state fMRI network between 55 brain regions where the top 20% of correlation are represented by a network connection (that means, the threshold was set so that the edge density was 20%). (C) Macaque cortical fiber-tract network of 95 brain regions. (D) Rat supragranular pyramidal cell neuronal network of layers II and III of the extrastriate visual cortex. Subplots C and D adapted from (Kaiser et al., 2009).

Establishing connections involves metabolic structural costs for building connections (especially for myelinated axons) as well as dynamic costs for transmitting action potentials. It is therefore natural to assume that these energy costs should be as low as possible (Cherniak, 1992; Chklovskii et al., 2002; Wen and Chklovskii, 2008). One possibility for reducing costs is to have a lower number of long-distance connections. The frequency of such long-distance connections, relative to short-distance connections, can be seen in the connection length histogram. It can also be tested how far away the combined length of all connections together, the total wiring length, is from the shortest-possible total wiring length. Such an optimal solution can be found in two ways: rearranging the connections of each node whereas the position of a node remains the same or rearranging the position of each node, swapping around the position of nodes, while retaining the connectivity of each node projecting to the same target nodes (not target positions).

Reducing the total length by reordering connections could lead to minimal wiring of a system. One possibility would be to establish connections ranked by their length; connecting nodes that are closest to each other first. However, this could result in a fragmented network where parts of the network are unreachable from many starting nodes. To secure reachability, start with a minimum spanning tree (cf. glossary) that connects $N$ nodes with $N$-1 edges and a minimal wiring length (Cormen et al., 2009), then add remaining connections again using short-distance connections as described before.

Alternatively, wiring length reductions in neural systems can be achieved by suitable spatial arrangement of the components. Under these circumstances, the connectivity patterns of neurons or regions remain unchanged maintaining their structural and functional connectivity, but the layout of components is perfected such that it leads to the most economical wiring. In the sense of this 'component placement optimization' (CPO; (Cherniak, 1994)), any rearrangement of the position of neural components, while keeping their connections unchanged, would lead to an increase of total wiring length in the network. However, studies on neuronal networks in *C. elegans* and on cortical networks in the macaque have shown that a reduction by 30-50% is possible (Kaiser and Hilgetag, 2006). For a small number of nodes, all possible arrangements of their positions can be tested. For larger networks, however, such an approach is not feasible: the number of possible layouts for $N$ nodes is $N$! (e.g. $10^{148}$ possibilities for 95 nodes). In those cases, optimal solutions can only be approximated through numerical routines such as simulated annealing (Metropolis et al., 1953) or others.

**6.2 Missing links: using spatial and topological features for network reconstruction**
As for other biological systems, incomplete data sets are a problem for brain connectivity studies. Are there ways to predict whether a connection between two nodes exist? One possibility, tested for the macaque fiber-tract and the *C. elegans* neuronal network, is to use local features of a pair of nodes to predict whether they are connected or not (Costa et al., 2007a). Topological features were node degree, clustering coefficient, characteristic path length, and Jaccard coefficient. Spatial or geometrical features included local density of nodes, coefficient of variation of the nearest distances, Cartesian coordinates of the nodes' center of mass, as well as the area size for nodes in the cortical network. Such an approach gave good estimates for reconstructing connections of the macaque visual cortex (Costa et al., 2007a). The prediction performance could be further improved through varying the contribution of each feature (training weights) (Nepusz et al., 2008).

**7. Conclusion**
This tutorial is a first introduction to connectome analysis looking at topological and spatial features of neural systems. There are several aspects of the connectome that are not covered here such as complexity (Tononi et al., 1994, 1996), divergence and convergence of information (Tononi and Sporns, 2003), and the comparison of types of connectivity (e.g. the link between structural and functional connectivity (Honey et al., 2009)). For the hierarchical organization, we only looked at topology but hierarchy also relates to the dynamics and spatial organization of neural systems (Hutt and Lesne, 2009; Jarvis et al., 2010; Kaiser and Hilgetag, 2010; Kiebel et al., 2009; Krumnack et al., 2010; Meunier et al., 2009; Rodrigues and Costa, 2009; Zamora-Lopez et al., 2010). We also only observed a snapshot of the connectome; however, neural systems change during individual development (Kaiser and Hilgetag, 2004b; Kaiser et al., 2009; Nisbach and Kaiser, 2007; van Ooyen, 2003; Van Ooyen, 2005), brain evolution (Ebbesson, 1980; Krubitzer and Kahn, 2003; Striedter, 2004), and throughout life through structural and functional plasticity (Butz et al., 2009; Friston et al., 1994; Hubel et al., 1977; Sur and Leamey, 2001). In addition, we can observe the dynamics in neural systems and simulate dynamics in network representations of the brain (Qubbaj and Jirsa, 2007).

In this work, we have outlined how routines from network analysis can be applied to neural systems. However, neural systems can also inform future work on the analysis of complex biological networks. Neural systems present several challenges for the field of network analysis. Neural systems differ from Erdös–Rényi and traditional small-world networks in that they are modular and hierarchical (Costa et al., 2011; Meunier et al., 2010; Sporns, 2011b). Network properties can be studied at different levels ranging from connectivity between brain areas, connectivity within areas, connectivity within columns (Binzegger et al., 2004), or connectivity of groups and ensembles. Another difference with respect to standard network models is that nodes, although treated as uniform at the global level of analysis, differ at the neuronal level in their response modality (excitatory or inhibitory), their functional pattern due to the morphology of the dendritic tree and properties of individual synapses, and their current threshold due to the history of previous excitation or inhibition. Such heterogeneous node properties can also be expected at the global level in terms of the size and layer architecture of cortical and subcortical regions. Theories where the properties and behavior of individual nodes differ, beyond their pattern of connectivity, are still rare. Another theoretical challenge is the comparison of network topologies and dynamics, e.g., between experiments and *in silico* studies (Izhikevich and Edelman, 2008; Kaiser et al., 2007a; Zhou et al., 2006).

In addition to theoretical challenges, network analysis also poses computational problems. The analysis of experimental network data, such as of correlation network between voxels for MRI recordings, can take a considerable amount of time. Whereas detecting all network motifs (Ribeiro et al., 2009) in a single 100-node correlation network is computationally feasible, long recordings could generate dozens of such correlation networks, which come with enormous computational demands. Similar demands arise from larger networks such as region of interest (ROI) networks with around 1,000 nodes. These problems also occur for electrophysiological recordings: Multi-electrode units with 4000 or more electrodes are now available. For the electrophysiology field, the CARMEN Neuroinformatics project (http://www.carmen.org.uk) addresses the storing, comparison, and analysis of large data sets; similar initiatives for neuroimaging are clearly needed and some projects along these directions already started. High-performance computing is also needed for large-scale simulations of neural circuits, such as the Blue Brain project for simulating activity within a single cortical column (Silberberg et al., 2005) or simulations of the whole brain (Izhikevich and Edelman, 2008).

In conclusion, connectome analysis provides several benefits to the field of neuroimaging. First, it gives a network representation of a human brain in that the size, shape, and position of brain regions are abstracted away. In this way, networks reduce the complexity of information yielded by neuroimaging recordings (Sporns, 2011a). Second, networks can be compared between humans. In particular, network analysis can identify differences between the brains of patients and control subjects. These changes can either be used for diagnosis of brain disorders or for evaluating treatment strategies. Third, connectomes, together with properties of individual nodes and edges as well as input patterns, form the structural correlate of brain function. Therefore, connectomes now increasingly form the basis of simulations of brain dynamics. These benefits of the emerging field of connectome analysis (DeFelipe, 2010) are now within reach due the availability of datasets, e.g. through the Human Connectome Project or the 1000 Functional Connectome Project (Biswal et al., 2010), the range of network analysis tools, and the computational feasibility of network analysis and simulation.

## 8. Acknowledgements
Part of this manuscript is based on lectures taught at Seoul National University and Newcastle University and I thank students and members of my labs for helpful comments. M.K. was supported by WCU program through the National Research Foundation of Korea funded by the Ministry of Education, Science and Technology (R32-10142), the Royal Society (RG/2006/R2), the CARMEN e-science project (www.carmen.org.uk) funded by EPSRC (EP/E002331/1), and (EP/G03950X/1).

## Appendix A
**Table A1**
Runtime complexity: Assessing the speed of network analysis algorithms.

Network features differ in the amount of time and computer memory it takes to calculate them. The runtime depends on the network feature, the implemented algorithm, the number of nodes $N$ and the number of edges $E$ of the network, and the speed of the processor. The memory size depends on the network representation and, in some cases, the organization of the algorithm. Using an adjacency matrix needs $N^2$ units of memory. Using adjacency lists saves memory using on the order of $E$ units but increases the runtime for several network measures when the edge density is very high.

Rather than using the actual calculation time, which strongly depends on processor speed, algorithms are categorized by the runtime complexity depending on the network size. This is called asymptotic analysis that looks at the growth of the running time instead of the absolute running time. For this, the asymptotic O-notation (pronounced: Big-Oh), meaning 'order-of', is used for getting information about the worst-case scenario of running time. The worst-case running time guarantees that an algorithm will not go above this upper limit. The O-notation shows the order of calculation steps that are performed by an algorithm. This means that only the largest terms that determine the runtime are kept:
$O(c N) = O(N)$          constants are neglected
$O(N^2 + N + c) = O(N^2)$     only the largest term of a polynomial is used
Using this notation, algorithms belong to complexity classes $P$, $NP$-hard, or $NP$-complete (Cormen et al., 2009).

Rather than looking at computational complexity theory, I will provide some examples for how long different network analysis algorithms take. Calculating the degree of a node takes $O(N)$ steps when an adjacency matrix is used and increases linearly with the number of nodes $N$. Note that for a directed network, it takes $2N$ steps for counting all incoming (column) and outgoing (row) connections, but the constant 2 is neglected in the O-Notation. Calculating the degree of all $N$ nodes takes $O(N^2)$ steps and calculating the shortest path between all pairs of nodes (characteristic path length) takes $O(N^3)$ steps; both being examples for polynomial increase with the number of nodes. Calculating the characteristic path is already a problem for large networks, but algorithms with non-polynomial runtime complexity are even worse: testing whether two graphs are identical, a graph similarity problem that occurs in motif analysis, takes $N$ factorial, $O(N!)$, steps whereas the travelling salesman problem of finding the shortest metric path to visit N cities takes $O(N^N)$ steps.

For large networks, some measures will take too long to calculate leaving three options: using a different network measure where calculations are computationally feasible, using parallel computing which works for some network measures, or applying the measure to a subset (sample) of the network. An example for sampling is the estimation of the characteristic path length of the world-wide-web with $8 \times 10^8$ nodes based on a sample of a few thousand nodes (Albert et al., 1999). Note, however, that sampling can become inaccurate especially when testing for power-law degree distributions (Stumpf et al., 2005).

**Table A2**
Glossary of network analysis terms.

**Adjacency (connection) matrix:** The adjacency matrix of a *graph* is an $n \times n$ matrix with entries $a_{ij} = 1$ if node *j* connects to node *i*, and $a_{ij} = 0$ if there is no connection from node *i* to node *j*. Sometimes, non-zero entries indicate the strength of a connection using ordinal scales for fiber strength or metric scales [-1; 1] for correlation networks.

**Average Shortest Path:** The average shortest path *ASP* is the global mean of the entries of the *distance matrix*. Normally, infinite values (non-existing paths) and values across the diagonal (loops) are not taken into account for the calculation of the mean of the distance matrix.

**Adjacency list:** List where each line represents one edge with information about the source node, the target node, and (optionally) the weight of the edge connecting both nodes.

**Betweenness centrality:** what proportion of all *shortest paths* are going through a node or edge of the network. These values are then called node betweenness or edge betweenness, respectively. See (Brandes and Erlebach, 2005) for other measures of centrality or influence.

**Characteristic path length:** The characteristic path length *L* (also called "path length") is the median of the mean shortest path length of all individual nodes (Watts, 1999).

**Clustering coefficient:** The clustering coefficient $C_i$ of node *i* is the number of existing connections between the node's neighbors divided by all their possible connections. The clustering coefficient ranges between 0 and 1 and is typically averaged over all nodes of a *graph* to yield the graph's clustering coefficient *C*.

**Cycle:** A *path* that links a node to itself.

**Degree:** The degree *k* of a node is the sum of its incoming (afferent) and outgoing (efferent) connections. The number of afferent and efferent connections is also called the in-degree and out-degree, respectively.

**Degree distribution:** Probability distribution of the degrees of all nodes of the network.

**Degree sequence:** The *N*-tuple $(k_1, \ldots, k_N)$ with $k_i$ as degree of node *i* and $k_i \leq k_{i+1}$ is called degree sequence.

**Distance:** The distance between a source node *i* and a target node *j* is equal to the *path length* of the shortest path. To talk about spatial distance instead, use the term metric distance.

**Distance matrix:** The entries $d_{ij}$ of the distance matrix correspond to the *distance* between node *i* and *j*. If no path exists, $d_{ij} = \infty$.

**Graph:** Graphs are a set of *N* nodes (vertices, points, units) and *E* edges (connections, arcs). Graphs may be undirected (all connections are symmetrical) or directed. Because of the polarized nature of most neural connections, we focus on directed graphs, also called digraphs.

**Hub:** Node with a degree that is much higher than for other nodes of the network. In sparse networks, even the degree of a hub might be relatively low.

**Kronecker symbol $\delta$:** $\delta_{i,j} = 1$ for $i = j$ and $\delta_{i,j} = 0$ otherwise.

**Loop:** A connection of a node onto itself (in other words: a cycle of length 1).

**Minimum spanning tree:** A minimum spanning tree is a tree that connects all nodes of a network with a weight less than or equal to the weight of every other spanning tree. In this context, the weight is usually the total metric wiring length of the tree when the network is a *spatial graph*.

**Modular graph:** A network with multiple modules.

**Motif:** Sub-graph with a certain number of nodes (usually 3, 4, or 5) that occurs significantly more often in a given network than in rewired networks with the same degree distribution.

**Path:** A path is an ordered sequence of distinct connections and nodes, linking a source node *i* to a target node *j*. No connection or node is visited twice in a given path.

**Path length:** The length of a *path* is equal to the number of distinct connections.

**Random graph:** An Erdös-Rényi graph with uniform connection probabilities and a binomial degree distribution. All node degrees are close to the average degree ("single-scale").

**Scale-free graph:** Graph with a power-law degree distribution. "Scale-free" means that degrees are not grouped around one characteristic average degree (scale), but can spread over a very wide range of values, often spanning several orders of magnitude.

**Small-world graph:** A graph in which the *clustering coefficient* is much higher than in a comparable random network, but the *characteristic path length* remains about the same. The term "small-world" arose from the observation that any two persons can be linked over few intermediate acquaintances.

**Spatial graph** (or spatial network): A network where each node has a spatial position. Usually, nodes in neural networks can have two- or three-dimensional positions in metric Euclidean space.